\documentclass[compsoc,conference,a4paper,10pt,times]{IEEEtran}
\IEEEoverridecommandlockouts
\usepackage{cite}
\usepackage{amsmath,amssymb,amsfonts}
\usepackage{algorithmic}
\usepackage{graphicx}
\usepackage{textcomp}
\usepackage{bmpsize}
\usepackage{xcolor}
\usepackage{lipsum}
\usepackage{balance}
\usepackage{caption}
\usepackage{subcaption}

\usepackage{tcolorbox}

\usepackage{enumitem}
\usepackage{booktabs}   
\usepackage{array}      
\usepackage{tcolorbox}
\usepackage{tikz}

\newcommand{\sharif}[1]{\textcolor{blue}{[Sharif's - #1]}}

\newcommand{\subheading}[1]{\noindent{\textbf{#1}}}

\usepackage[colorlinks=true,urlcolor=black]{hyperref}
\def\BibTeX{{\rm B\kern-.05em{\sc i\kern-.025em b}\kern-.08em
    T\kern-.1667em\lower.7ex\hbox{E}\kern-.125emX}}
\begin{document}

\title{ThreatModeling-LLM: Automating Threat Modeling using Large Language Models for Banking System} 

\author{%
  \IEEEauthorblockN{%
    Tingmin Wu\IEEEauthorrefmark{1}\textsuperscript{\textsection},
    Shuiqiao Yang\IEEEauthorrefmark{1}\textsuperscript{\textsection},
    Shigang Liu\IEEEauthorrefmark{1},
    David Nguyen\IEEEauthorrefmark{1},
    Seung Jang\IEEEauthorrefmark{1} and
    Alsharif Abuadbba\IEEEauthorrefmark{1}%
  }%
  \IEEEauthorblockA{\IEEEauthorrefmark{1} CSIRO's Data61}%
}

\maketitle
\begingroup\renewcommand\thefootnote{\textsection}
\footnotetext{Equal contribution}
\endgroup

\begin{abstract}

Threat modeling is a crucial component of cybersecurity, particularly for industries such as banking, where the security of financial data is paramount. 
Traditional threat modeling approaches require expert intervention and manual effort, often leading to inefficiencies and human error. 
The advent of Large Language Models (LLMs) offers a promising avenue for automating these processes, enhancing both efficiency and efficacy. However, this transition is not straightforward due to three main challenges: (1) the lack of publicly available, domain-specific datasets, (2) the need for tailored models to handle complex banking system architectures, and (3) the requirement for real-time, adaptive mitigation strategies that align with compliance standards like NIST 800-53.

In this paper, we introduce ThreatModeling-LLM, a novel and adaptable framework that automates threat modeling for banking systems using LLMs.
ThreatModeling-LLM operates in three stages: 1) dataset creation, 2) prompt engineering and 3) model fine-tuning. 
We first generate a benchmark dataset using Microsoft Threat Modeling Tool (TMT). Then, we apply Chain of Thought (CoT) and Optimization by PROmpting (OPRO) on the pre-trained LLMs to optimize the initial prompt. 
Lastly,  we fine-tune the LLM using Low-Rank Adaptation (LoRA) based on the benchmark dataset and the optimized prompt to improve the threat identification and mitigation generation capabilities of pre-trained LLMs. 
The experimental results demonstrate that our proposed scheme substantial improvements over the pre-trained LLMs, significantly enhancing the model's ability to identify threats and suggest mitigations. For example, the accuracy of identifying mitigation codes improves from 0.36 to 0.69 on Llama-3.1-8B-Instruct (short for Llama-3.1-8B).
The results illustrate that the combination of prompt engineering and fine-tuning techniques is highly effective for automated threat modeling, making ThreatModeling-LLM a robust and flexible solution for real-world applications in banking and beyond.
\end{abstract}


\section{Introduction}



\begin{figure}[t]  
    \includegraphics[width=0.49\textwidth]{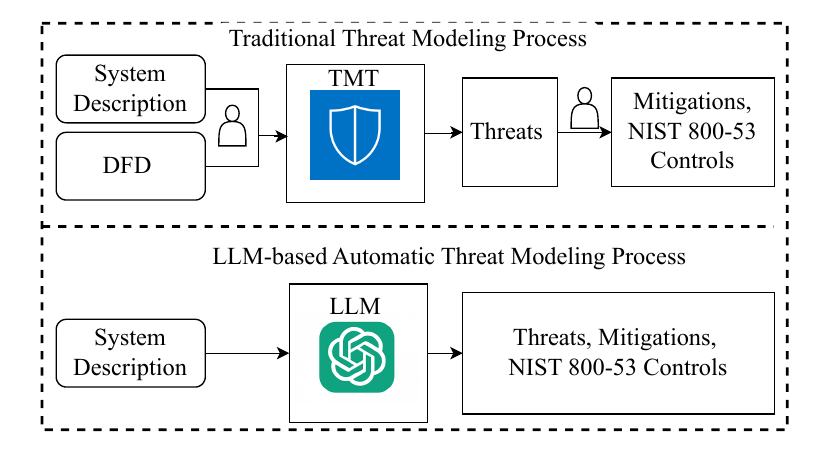}
    \caption{Comparison of Traditional method and LLM-based method. The traditional method (top)
    requires manual creation of Data Flow Diagrams (DFDs). After threats are identified, additional manual effort is needed to map them to mitigations and code. In contrast, the LLM-based process (bottom) streamlines the workflow by using system descriptions as input to automatically generate threats, corresponding mitigations, and the
    NIST 800-53 controls.}
    \label{fig:Human_threat_Modeling}
\end{figure}


Threat modeling is a critical cybersecurity process that identifies potential threats and suggests mitigations for system designs using frameworks like Microsoft’s STRIDE~\cite{xiong2022cyber}. It plays a vital role in proactively addressing vulnerabilities and preventing security breaches, which can lead to significant financial and reputational damages~\cite{ibrahim2020challenges}. For instance, threat modeling can block intrusion attempts and prevent hijacking of privileged accounts, significantly reducing risks in critical systems~\cite{stevens2018battle}. However, the traditional approach is labor-intensive, requiring manual efforts for Data Flow Diagram (DFD) creation, threat identification, and mapping to mitigations, which makes it inefficient and prone to human error.  Figure~\ref{fig:Human_threat_Modeling} shows the traditional process relies heavily on tools like the Microsoft Threat Modeling Tool (TMT), which demands extensive manual input at each stage. This is particularly challenging in dynamic sectors like banking, where the rapid evolution of online services and increasing sophistication of threats have intensified the need for more efficient, automated threat modeling solutions~\cite{crothers2023machine}. Traditional methods struggle to keep up with the complexity of confidentiality, integrity, and privacy requirements in banking systems, underscoring the urgency for automation. 

Large Language Models (LLMs) such as GPT-3~\cite{brown2020language} and Llama-3~\cite{dubey2024llama} offer promising potential to transform the threat modeling landscape. Traditional methods, such as pytm~\cite{owasp_pytm}, do not provide direct mappings to NIST 800-53 standards, which are critical for compliance and comprehensive security analyses. LLMs can process textual descriptions of system designs, automatically identifying threats and suggesting corresponding mitigations. This shift not only accelerates the process but also enhances accuracy by reducing manual intervention. For instance, commonly utilized industry tools like STRIDEGPT~\cite{stride_gpt} and Cyber Sentinel~\cite{kaheh2023cyber} show the trade-offs between automation and precision; STRIDEGPT, while automating threat identification, produces unstable results. Cyber Sentinel, despite its adaptability to new threats, offers limited mitigation strategies. These examples underscore a prevalent trade-off between specializes capabilities and comprehensive functionality across these tools. 
While pre-trained LLMs, have demonstrated impressive results in various NLP tasks, directly applying them to threat modeling in banking systems is insufficient. Pre-trained LLMs lack domain-specific knowledge and struggle to understand complex banking architectures, resulting in inconsistent threat identification and  mitigation suggestions. Moreover, they are not specifically designed to generate mitigation codes aligned with compliance standards like NIST 800-53, which is essential for banking security. These drawbacks highlight that without additional adaptation, LLMs fall short of meeting the precision, compliance, and context-specific needs of threat modeling in the banking sector.
%
However, the adaptation of LLM for threat modeling is non-trivial and poses several challenges:




\subheading{Challenge 1: Lack of publicly available datasets}. A significant challenge in threat modeling analysis is the lack of publicly available datasets, especially for complex systems like banking. Traditionally, researchers and security experts generally manually assess potential threats within DFDs, a process that is not only labor-intensive but also prone to human error. To transition towards a data-driven approach for automatically identifying threats in banking system-based DFDs, real-world datasets are essential. These datasets provide crucial information about threats and mitigations in practical scenarios, serving as the foundation for building and training automated tools that can efficiently detect and address security vulnerabilities.
However, \textit{the creeation of a real-world dataset for automatic threat modeling remains a challenging problem in the field.} 
    
\subheading{Challenge 2: Tailored LLMs for banking systems}. While LLMs have achieved remarkable success in fields like natural language processing, software security, and network security, applying them to threat analysis in banking systems is an underexplored area. The unique structure and operational complexity of banking systems require specialized threat models that can understand the specific vulnerabilities in financial transactions, user authentication, and data flow between systems. This gap hinders the efficient identification and modeling of threats unique to banking infrastructures. \textit{Developing an efficient and effective LLM-based system for bank system-based threat analysis poses an important and creative research question that remains unsolved.}
    
\subheading{Challenge 3: Lacking an automatic mitigation strategies}: Once threats are identified, the next critical step is developing effective, real-time mitigation to safeguard the system. However, this is a complex task due to the dynamic and evolving nature of threats within financial systems. Developing novel mitigation strategies requires deep expertise in both banking operations and security protocols, as well as sophisticated algorithms that can adapt to changing threats. Automating this process is particularly challenging because it demands solutions that can respond to threats in real-time while maintaining system efficiency and compliance with stringent banking regulations. Therefore, \textit{creating an automatic mitigation strategy remains a challenge that still needs to be addressed to ensure continuous improvement in system security and the protection of sensitive financial data.}

To address the first challenge, we created \textbf{the first benchmark dataset} in the community by designing various types of banking systems. For each system, we used the TMT to draw the DFDs based on the application design documents. The TMT-generated threats and human-annotated mitigation strategies using the NIST 800-53 served as the ground truth for fine-tuning the LLMs, ensuring that the dataset accurately reflected real-world security scenarios.
For the \textbf{second and third challenges}, we propose combining prompt engineering and fine-tuning methods to create a customized LLM model, ThreatModeling-LLM, specifically for identifying banking system threats and mitigations.
\subheading{Prompt Engineering}.
We explore different prompt templates to find the optimal structure for the LLM to produce accurate threat and mitigation outputs. Chain-of-Thought (CoT)~\cite{wei2022chain} is used to make the model explicitly reason through intermediate steps, and OPRO (Optimization by PROmpting)~\cite{yang2024large} is applied to refine the prompts iteratively. These techniques help improve the quality of generated responses when identifying threats and suggesting mitigations.
%
%
\begin{table*}[!t]
\centering
\caption{Summary of  Related Works in Threat Modeling.}
\label{tab:related_work_summary}
\begin{tabular}{|p{2.5cm}|p{3cm}|p{2cm}|p{3.5cm}|p{3.5cm}|}
\hline
\textbf{Study} & \textbf{Approach} & \textbf{Domain Type} & \textbf{Strengths} & \textbf{Limitations / Gaps} \\ \hline
Tong and Ban~\cite{xin2014online} & STRIDE + Threat Tree Analysis & Banking & Enhanced threat analysis efficiency & Lacks automation; high manual effort \\ \hline
Chattopadhyay and Sripada~\cite{chattopadhyay2023security} & Threat Modeling Framework for Mobile Banking & Banking & Comprehensive threat detection & Lacks specific mitigations \\ \hline
M{\"o}ckel and Abdallah~\cite{mockel2010threat, mockel2011understanding} & Threat Modeling in SDL for E-Banking & Banking & Early-stage threat identification & High manual effort \\ \hline
Hassan et al.~\cite{bakar2021iot} & IoT Security Risks in Banking & Banking & Proposes blockchain-based measures & No automated threat modeling mechanism \\ \hline
Aijaz et al.~\cite{aijaz2023threat} & TMA for Healthcare-IT Systems & Healthcare & Improved threat identification & No real-time mitigation; lacks NIST compliance \\ \hline
Beozzo~\cite{beozzo2023modern} & Agile Threat Modeling & Corporate/Agile & Scalable, governance-focused approach & Not adapted to banking-specific needs \\ \hline
Abuabed et al.~\cite{abuabed2023stride} & Cybersecurity Framework for Automobiles & Automotive & Integrates STRIDE and CVSS & Struggles with evolving banking threats \\ \hline
Ananthapadmanabhan and Achuthan~\cite{ananthapadmanabhan2022threat} & Threat Intelligence for Cloud Systems & Cloud/IT & Effective attacker behavior capture & No fine-tuning for domain-specific detection \\ \hline
Rose et al.~\cite{de2022threma} & ThreMA (Ontology-Based) & Public Sector & Automated threat identification & Static framework; limited banking adaptation \\ \hline
Schaad and Reski~\cite{reski2019open} & OVVL Framework & Software Development & Early-stage threat modeling & Lacks domain-specific focus and compliance alignment \\ \hline
\end{tabular}
\end{table*}
\subheading{Model Fine-Tuning}. Based on the generated prompt, we fine-tune a base LLM (such as Llama-3.1-8B) using our created dataset to empower the LLMs with the abilities to generate more accurate threats and mitigations based on the text input.  
The fine-tuning process involves Low-Rank Adaptation (LoRA) \cite{hu2021lora}, which allows efficient adaptation of the model to domain-specific tasks like threat identification in banking systems. The fine-tuning data includes DFD descriptions, identified threats, mitigations, and their respective NIST 800-53 control codes. Fine-tuning enables the model to grasp the unique vulnerabilities and mitigations needed in financial systems.

Our experiments show that combining prompt engineering and fine-tuning outperforms using the either technique alone. ThreatModeling-LLM demonstrates significant improvements in the accuracy of threat identification and mitigation generation, providing a more effective and automated threat modeling process tailored for the banking sector.
Our key contributions are as follows:

\begin{itemize}

\item We introduce the innovative ThreatModeling-LLM framework, specifically designed to automate threat modeling for banking systems. This framework is uniquely tailored to address the complexities of the banking domain, combining advanced prompt engineering with fine-tuning techniques to to enhance the capability of pre-trained LLMs. Specifically we optimize the prompt by CoT and OPRO. We then fine-tuning the model with our specialized dataset using the optimized prompt.  These techniques encourage the model to explicitly reason through intermediate steps and iteratively refine its responses, leading to more accurate and detailed identification of threats and generation of mitigation strategies aligned with NIST 800-53 control codes.

\item To overcome the scarcity of publicly available datasets for banking threat modeling, we have meticulously designed a specialized training dataset that features 50 different banking system applications and use cases. This involved creating various types of banking system scenarios and using the TMT to draw DFDs based on application design documents. The TMT-generated threats and human-annotated mitigation strategies serve as ground truth, ensuring the dataset reflects real-world security concerns. This dataset is essential for effectively fine-tuning the LLM to understand and identify banking-specific threats and mitigation.

\item We demonstrate that combining fine-tuning with advanced prompt engineering techniques significantly improves the LLM's performance in threat modeling tasks.  Specifically, for the Llama-3.1-8B model, the synergy of prompt engineering and fine-tuning increased accuracy from 0.36 to 0.69, precision from 0.49 to 0.73, recall from 0.36 to 0.73, and text similarity from 0.944 to 0.9792. This marked improvement underscores the effectiveness of our methods in sharpening the model’s ability to accurately identify threats and mitigations, making it a potent tool for cybersecurity in the banking sector.


\end{itemize}



\section{Related Work}


Automated threat modeling has evolved through both traditional and AI-based
approaches, aiming to enhance cybersecurity across various domains. In this section, we review notable works in threat modeling for banking systems and other
domains, identifying gaps that motivate the development of ThreatModeling-LLM.

\textbf{ThreatModeling for Banking Systems:}
Tong and Ban~\cite{xin2014online} combined STRIDE with threat tree analysis to improve threat analysis efficiency in online banking. This hybrid approach provided deeper insights into security risks but lacked automation, making it labor-intensive. 
Chattopadhyay and Sripada~\cite{chattopadhyay2023security} reviewed major threats to mobile banking, offering a comprehensive framework for detection and mitigation, yet the proposed solution was limited to identifying broad categories without generating specific mitigations. 
M{\"o}ckel and Abdallah~\cite{mockel2010threat, mockel2011understanding} emphasized integrating security into the software development lifecycle (SDL) for e-banking, advocating early-stage threat modeling using tools like Microsoft SDL. Despite highlighting the importance of threat modeling in design, the manual nature of their approach still required significant human input. Hassan et al.~\cite{bakar2021iot} focused on IoT's impact on banking, proposing blockchain-based measures for IoT security risks, but lacked an automated threat modeling mechanism that aligns with compliance standards like NIST 800-53.


\textbf{Threat Modeling for Other Domains}: Several studies have explored automated threat modeling beyond banking system.
Aijaz et al.~\cite{aijaz2023threat} introduced Threat Modeling and Analysis (TMA) for healthcare-IT, focusing on system vulnerabilities and attacker behavior. Although TMA improved threat identification, it did not offer real-time, adaptive mitigation strategies. Beozzo~\cite{beozzo2023modern} proposed a novel approach for Agile corporate environments, addressing scalability and governance but not covering domain-specific needs like banking compliance.

Abuabed et al.~\cite{abuabed2023stride} tailored a cybersecurity analysis framework for modern automobiles, integrating STRIDE, Attack Tree Analysis, and CVSS. However, their method struggled with the complexity of identifying threats across evolving banking infrastructures. Ananthapadmanabhan and Achuthan~\cite{ananthapadmanabhan2022threat} integrated threat modeling with threat intelligence in cloud systems using Splunk, but this approach lacked fine-tuning for domain-specific threat detection. Rose et al.~\cite{de2022threma} introduced ThreMA, an ontology-driven threat modeling tool, while Schaad and Reski~\cite{reski2019open} proposed OVVL for early-stage threat identification. Both studies advanced automation but were limited by static frameworks that did not adapt to specific banking architectures.

\textbf{Summary and Identified Gaps}: Table~\ref{tab:related_work_summary} summarizes key related works, their strengths, and limitations. While existing approaches contribute to various aspects of threat modeling, they often lack automation, domain-specific adaptation, or compliance with standards like NIST 800-53. These gaps highlight the need for a more flexible and adaptable approach, motivating the development of \textbf{ThreatModeling-LLM}.


\section{Preliminaries, Problem Definition and Motivation}

%
\subsection{Preliminaries}




\begin{table*}[!t]
\centering
\caption{Description of STRIDE Framework Threats and Desired Properties. Source: \url{https://en.wikipedia.org/wiki/STRIDE_model}}
\label{tab:stride}
\begin{tabular}{|c|c|p{10cm}|}
\hline
\textbf{Threat} & \textbf{Desired Property} & \textbf{Threat Definition} \\ \hline
Spoofing & Authenticity & Pretending to be something or someone other than yourself \\ \hline
Tampering & Integrity & Modifying something on disk, network, memory, or elsewhere \\ \hline
Repudiation & Non-repudiability & Claiming that you didn't do something or were not responsible; can be honest or false \\ \hline
Information Disclosure & Confidentiality & Someone obtaining information they are not authorized to access \\ \hline
Denial of Service & Availability & Exhausting resources needed to provide service \\ \hline
Elevation of Privilege & Authorization & Allowing someone to do something they are not authorized to do \\ \hline
\end{tabular}
\end{table*}

\textbf{Microsoft STRIDE} \cite{shevchenko2018threat} is a framework used for threat modeling in software security.
As shown in Table \ref{tab:stride}, it stands for Spoofing, Tampering, Repudiation, Information Disclosure, Denial of Service, and Elevation of Privilege, representing the various categories of security threats that need to be addressed. When using Microsoft STRIDE for threat modeling analysis, the process begins with the identification and categorization of potential security threats to a system. Each category corresponds to a specific kind of threat. For example, `Spoofing' involves an unauthorized user impersonating another to gain access to a system, while `Tampering' refers to the unauthorized modification of data.

By applying the STRIDE framework, security analysts systematically explore each threat category in relation to the target system. They assess the system's architecture to identify where and how these threats could potentially be realized. This involves examining data flows, authentication mechanisms, network interfaces, and other relevant aspects of the system. Once threats are identified, they are documented, and the system's vulnerabilities that could be exploited by these threats are pinpointed.

The final phase of the Stride methodology involves proposing and prioritizing mitigations for identified threats. This may include implementing secure coding practices, enhancing authentication protocols, applying encryption, or introducing intrusion detection systems. Stride not only helps in recognizing potential threats but also plays a crucial role in the design phase by guiding developers to integrate security measures early in the development process, thus reinforcing the system's defense against malicious attacks and reducing the risk of security breaches.

\begin{table*}[!t]
\centering
\caption{Existing Automated Threat Modeling Explanations.}
\label{table:tool_comparsion}
\begin{tabular}{|>{\raggedright}p{2cm}|>{\raggedright\arraybackslash}p{5.5cm}|>{\raggedright\arraybackslash}p{5.5cm}|}
\hline
\textbf{Model} & \textbf{Strength} & \textbf{Weakness} \\ \hline
pytm & Integration with Python codebases & Lacks direct NIST 800-53 mapping \\ \hline
STRIDEGPT & Applies STRIDE in an automated manner & No consistent categorization, unstable results \\\hline
Cyber Sentinel & High adaptability to new threats & Limited mitigation capabilities \\\hline
Raw LLM (ChatGPT) & Broad contextual knowledge & Inconsistent mitigation suggestions, lacks deep technical mappings \\ \hline
\end{tabular}
\end{table*}

\textbf{NIST Cybersecurity Framework} is developed by National Institute of Standards and Technology (NIST) to provide high-level cybersecurity outcomes for different sizes of business. 
The framework has five functions to organize the cybersecurity activities, including Identify, Protect, Detect, Respond and Recover. Identify represents understanding cybersecurity risks of the relevant organization, then Protect develops effective security protective process to maintain the running of important services. Detect refers to the discovery of the cybersecurity events, and Respond takes actions to the incident. Lastly, Recover supports the recovery to the service and minimize the impact of the cybersecurity incidents.
\textbf{NIST 800-53}\footnote{Source: https://csrc.nist.gov/pubs/sp/800/53/r5/upd1/final} complements the framework by offering a detailed catalog of security and privacy controls. While the framework outlines the ``what'' of cybersecurity (the essential functions), NIST 800-53 provides the ``how'' by specifying technical and organizational safeguards. These controls can be mapped to the framework’s functions, making NIST 800-53 an operational tool to achieve the cybersecurity outcomes outlined in the framework. The controls are categorized into 20 families, such as AC (Access Control) and IR (Incident Response), tailored to enhance security across various organizational environments.

\subsection{Problem Definition}
In this section, we formally define the research problem.
As shown in Figure~\ref{fig:Human_threat_Modeling}, traditional threat modeling techniques, such as those based on Microsoft’s STRIDE framework, rely heavily on manual analysis, which is time-consuming and prone to human error. Furthermore, the process of identifying appropriate mitigation strategies and ensuring compliance with standards like NIST 800-53 is complex and resource-intensive. In this work, we aim to automate the threat modeling process for banking systems by leveraging LLMs as illustrated in Figure \ref{fig:Human_threat_Modeling}. 
Given a textual description of a banking system design, such as an ATM system, the goal is to automatically identify potential security threats based on the STRIDE framework and generate the mitigation strategies based on NIST 800-53. 

Let $X$ represent the designing document of a banking system.  This input \( X \) is a sequence of words (tokens), where:
\[
X = [x_1, x_2, \dots, x_n]
\]
where \( x_i \) represents the \( i \)-th token in the text, and \( n \) is the total number of tokens in the input description. The goal is to transform this input into two outputs:
\begin{itemize}
    \item Threats Identification: \( T \)
    \item Mitigation Strategy: \( M \)
\end{itemize}

\subsubsection*{Threats Identification}
Let \( T = \{t_1, t_2, \dots, t_k\} \) represent the set of identified threats, where \( t_i \) corresponds to the \( i \)-th threat. Each threat \( t_i \) is a function of the input description \( X \) and is defined based on the STRIDE framework categories (Spoofing, Tampering, Repudiation, Information Disclosure, Denial of Service, Elevation of Privilege):
\[
t_i = f_{\text{STRIDE}}(X), \quad \forall t_i \in \{S, T, R, I, D, E\}
\]
where \( S, T, R, I, D, E \) represent the STRIDE categories. The function \( f_{\text{STRIDE}} \) maps the input \( X \) to one or more threat categories based on the analysis of the data flow diagram (DFD).

\subsubsection*{Mitigation Strategy}
Let \( M = \{m_1, m_2, \dots, m_k\} \) represent the set of mitigation strategies, where \( m_i \) corresponds to the \( i \)-th mitigation strategy associated with threat \( t_i \). Each mitigation strategy \( m_i \) is a function of both the identified threat \( t_i \) and the control codes defined by the NIST 800-53 standard. The mitigation strategy \( m_i \) is given by:
$$
m_i = g_{\text{NIST}}(t_i), \quad \forall m_i \in \{NIST\ 800-53\ \text{Control Codes}\}
$$

where \( g_{\text{NIST}} \) maps the identified threat \( t_i \) to the corresponding NIST 800-53 mitigation control code.

\subsubsection*{Objective}
The system’s objective is to generate the set of pairs \( \{(t_i, m_i)\}_{i=1}^k \) from the input description \( X \) using the fine-tuned LLM model \( h \), where each pair \( (t_i, m_i) \) corresponds to an identified threat and its respective mitigation strategy based on the STRIDE framework and NIST 800-53 control codes.

The overall transformation can be formalized as:

\[
\{(t_i, m_i)\}_{i=1}^k = h(X)
\]

where \( h(X) \) is the function representing the LLM-based process that identifies the threats \( t_i \) and generates the corresponding mitigations \( m_i \) based on NIST 800-53 standard.

    

\begin{figure*}[!t]
    \centering
    \includegraphics[width=0.9\linewidth]{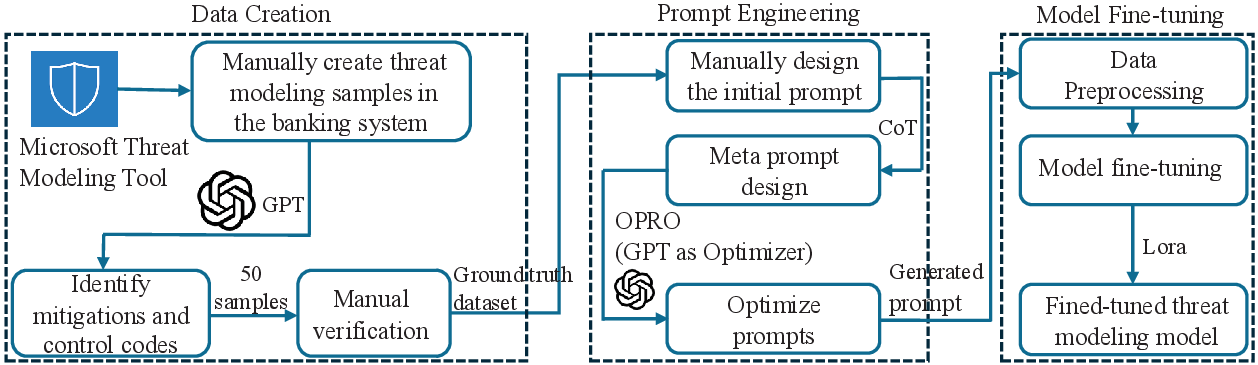}
    \caption{System Overview of ThreatModeling-LLM: (i) Data Creation: Utilizes the Microsoft Threat Modeling Tool to manually generate threat modeling samples, comprising 50 samples verified manually to construct a ground truth dataset. (ii) Prompt Engineering: Involves manually designing the initial prompt for a Large Language Model (LLM), followed by optimizing these prompts to enhance model responses. (iii) Model Fine-tuning: This phase includes the fine-tuning of the threat modeling model using the LLM to improve its accuracy and reliability in threat detection, and mitigation generation (i.e., NIST 800-53 control codes). 
}
    \label{fig:overview}
\end{figure*}

\subsection{Motivation}
\label{sec:motivation}

To investigate the state-of-the-art, we examined four notable automated threat modeling tools, encompassing both industry-standard and emergent GPT-based technologies. These tools are Cyber Sentinel (CS)~\cite{kaheh2023cyber}, pytm~\cite{owasp_pytm}, STRIDEGPT (SG)~\cite{stride_gpt}, and Raw LLM (RL) (using ChatGPT)~\cite{chatgpt} stand out for their unique capabilities. Cyber Sentinel is renowned for its adaptability to new threats, providing proactive security measures. pytm, specifically designed for Python applications, integrates threat modeling directly into the development process, facilitating seamless security assessments. STRIDEGPT leverages the STRIDE methodology through automation to efficiently pinpoint potential threats. Lastly, Raw LLM (ChatGPT) offers broad contextual knowledge, making it a versatile tool for general threat analysis.

We summarize the strengths and weaknesses of the tools in Table~\ref{table:tool_comparsion}. The limitations of the four automated threat modeling tools, Cyber Sentinel, pytm, STRIDEGPT, and Raw LLM (ChatGPT), highlight the challenges in balancing strengths with functional shortcomings within the realm of cybersecurity modeling. Cyber Sentinel, while highly adaptable to new threats, suffers from limited capabilities in offering specific mitigation strategies, which restricts its utility in proactive threat management. pytm, despite its seamless integration with Python environments, does not provide direct mappings to NIST 800-53 standards, which are critical for compliance and comprehensive security analyses. STRIDEGPT, which leverages the STRIDE methodology to automate threat identification, struggles with consistent categorization and produces unstable results, undermining its reliability. Lastly, Raw LLM (ChatGPT) offers expansive contextual knowledge but falls short in providing consistent, technically precise mitigation suggestions and lacks the capability to deeply map technical controls, which are essential for detailed threat remediation and control implementation. These limitations underscore a prevalent trade-off between specializes capabilities and comprehensive functionality across these tools, signaling the need for further refinement and development to enhance their applicability and effectiveness in diverse security scenarios.


\section{The Proposed ThreatModeling-LLM}

\subsection{Overview}
The comprehensive system overview is depicted in Figure~\ref{fig:overview}. Our system consists of three core components that streamline threat modeling. The first component, Data Creation, utilizes the Microsoft Threat Modeling Tool (TMT) to manually generate samples for threat modeling, with 50 samples manually verified to establish a ground truth dataset. The second component, Prompt Engineering, involves the manual design of initial prompt for an LLM, which are then optimized to improve the model’s response effectiveness. The final component, Model Fine-tuning, focuses on refining the threat modeling model through precise LLM fine-tuning, ensuring high accuracy and reliability in the bank applications.

\subsection{Dataset Creation and Verification}
To study automatic threat modeling using LLM in banking systems, one needs to prepare a well-organized dataset. However, as far as we know, there is no publicly available dataset. Therefore, we need to prepare the dataset, the dataset should meet the following requirements: 1) it should reflect real-world scenarios; 2) it should be generated using a publicly available and widely used tool in the security community; 3) it should include threats, mitigation, and the related control code. 

\begin{figure*}[!t]
    \centering
    \includegraphics[width=1\linewidth]{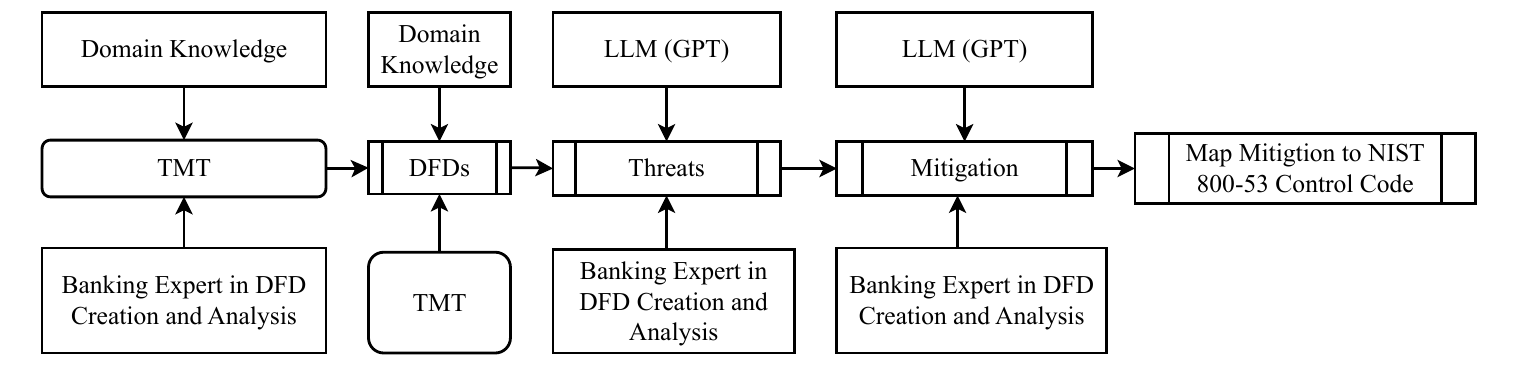}
    \caption{Dataset Creation Framework.}
    \label{fig:data_flow_framework}
\end{figure*}

In light of these requirements, this work first uses the TMT to generate different DFDs for banking systems. TMT is a core part of the Security Development Lifecycle, enabling users to create data flow diagrams and identify potential threats early. Designed for non-security experts, it simplifies threat modeling by providing standard visual notations and guidance, helping to mitigate security risks in software development. We use the Windows System. The windows operating system is Microsoft Windows Server 2019 Standard with 8GB RAM. Then, we use TMT to identify the related threats. 
After that, we employ LLM (i.e. GPT) to automatically identify the mitigations, and also manually check all the mitigation with local banking experts in DFD analysis. It is worth noting that, in the processes of mitigation generation and the map of mitigation to the NIST 800-53 control code generation, we have been working closely with our collaborator from the local bank, and the security expert, along with our security expert, has manually checked all the mitigations to ensure there is no noise. After this, we also employ LLM to map all the mitigations to the NIST 800-53 control codes, as our collaborator requires these codes to address all possible threats. We also want to point out that human experts have been involved in this process to ensure the quality of the mapping and that no noise is introduced. Figure \ref{fig:data_flow_framework} shows the processes for the dataset generation process.

\begin{figure}[!t]
    \centering
    \includegraphics[width=1\linewidth]{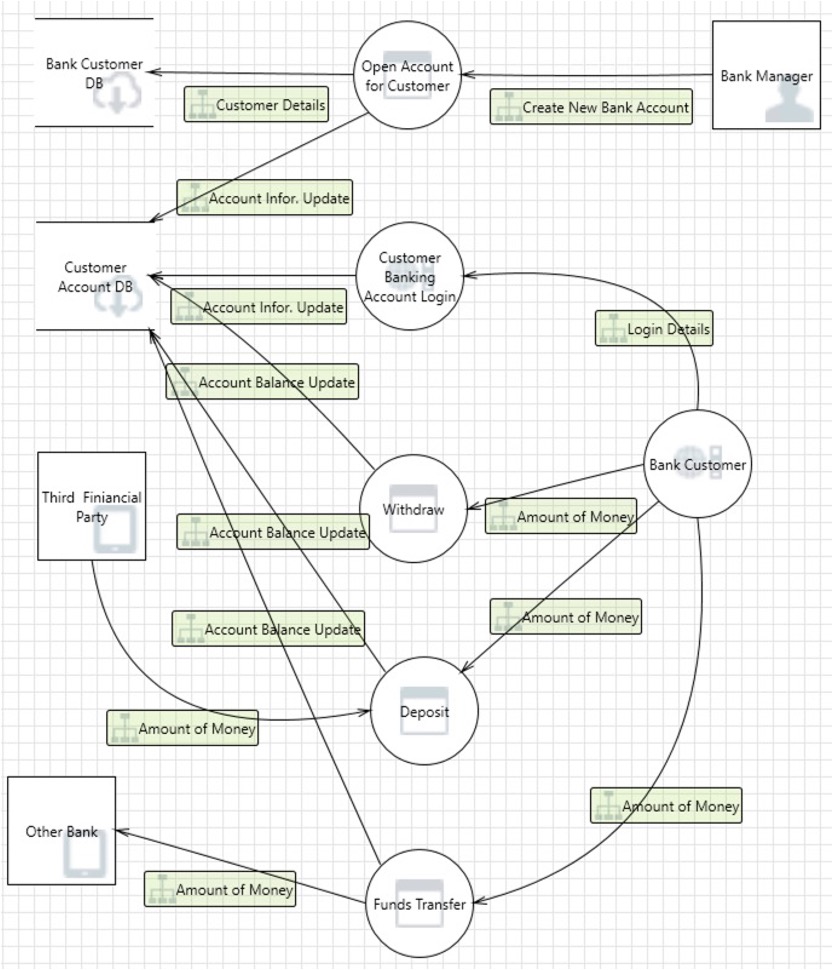}
    \caption{A light example of the Generated DFD.}
    \label{fig:example_DFD}
\end{figure}

Specifically, for the first step, we work closely with our local bank collaborator in preparing the DFD. For example,  for the ATM DFD, we first identify all the External entities including customer client; processes such as Manage Bank Customer Information, Bank Customer Information Management, Account Information Update and so on; data stores such as Bank Customer Database and transaction record; data flows including Transaction Request, Confirmation, Cash Out \& Receipts and so on; the relationship between the element such as Customers interact with the Manage Bank Customer Information process, which accesses the Bank Customer Database, Customers initiate a Transaction Request, which is handled by the Bank Customer Information Management process. Based on this information, we use TMT to draw the DFD in both the default setting and the managed setting as well. For the managed setting, the parameters are: the condition `Running as' will be changed from `no' to `network service', the `Isolation level' will be changed from  `no' to `AppContainer', the `Accepts Input From' will be changed from  `no' to `Kernel, System, or Local Admin' and so on, we will make all the setting publicly available along with our dataset. For more information
please refer to \url{anonymous}. Figure \ref{fig:example_DFD} provides an example of a DFD which is a Bank Account DFD. One can see, it includes: 1) External Entities such as bank customer, third financial party, other bank etc.; 2) Data Store such as customer account DB; 3) Processes such as open account for customer, customer banking account login etc. 

Once the DFD is ready, we will use TMT to produce the possible threats and save then into files. Afterward, we need to prepare mitigations based on the identified threats. To achieve this, we employ LLMs (e.g., GPT) to automatically identify the mitigations.
To ensure the generated mitigations are practical and applicable to real-world scenarios, we manually verify them with local banking experts in DFD analysis. For example, when facing a data tampering threats, there can be different mitigations such as SC-7 or SC-8. Actually, both SC-7 and SC-8 are critical for protecting the communications and system interfaces from potential threats such as unauthorized access, and data tampering, they are controls related to system and communications protection. In this case, we will manually check whether the system need a Boundary Protection (SC-7) or Transmission Confidentiality and Integrity (SC-8). The final version of the mitigations, after manual review, will be used for the next step.


For the third step, we use GPT-3.5 to map the mitigation to the NIST 800-53 control code. As a result, the prediction will include the threats, mitigations, and corresponding control codes. For example, if the identified threat category is Spoofing, the threat might be: `An attacker could impersonate a bank customer or IoT device to gain unauthorized access to the system. Spoofing can occur at the Web Service or IoT Device level.' The mitigations could be: `Implement strong authentication mechanisms, such as multi-factor authentication (MFA) for bank customers and secure authentication for IoT devices (e.g., certificates)' and `Use secure communication channels (e.g., TLS/SSL) to prevent identity spoofing.' The related NIST 800-53 control codes would be `IA-2: Identification and Authentication (Organizational Users)' and `SC-12: Cryptographic Key Establishment and Management.

All in all, by following these steps, we created 50 DFD samples. For each sample, we will prepare a file that includes the description of the DFD, the threats, the mitigation, and the mapping of NIST 800-53 control codes. These samples will be used for fine-tuning the LLM (based on 40 samples) and evaluating the LLM models (based on 10 samples).

\subsection{Prompt Engineering}
Prompt engineering represents a critical aspect of leveraging language models effectively, acting as the interface between human intentions and machine understanding. Prompt engineering involves crafting inputs that guide AI models, particularly LLMs, to generate desired outputs with higher precision and relevance. This discipline has become increasingly significant with the rise of more sophisticated AI models that are capable of understanding and generating human-like text.

Traditional prompt engineering has evolved from simple iterative refinements to adopting more systematic approaches. One prominent method within this evolution is the CoT prompting~\cite{wei2022chain}. This technique involves instructing the model to verbalize its intermediate reasoning steps or cognitive processes as it approaches a solution. By simulating a more transparent thought process, CoT helps align the model’s responses more accurately with complex problem-solving tasks, significantly improving the output’s clarity and correctness.

Recent prompt engineering has focuses more on dynamic generation of new prompts. Another innovative technique is OPRO (Optimization by PROmpting)~\cite{yang2024large}, which focuses on the dynamic refinement of prompts in response to evolving dialogue contexts and the model’s prior outputs. This adaptability is especially valuable in interactive settings where user queries can progressively modify the scope or specificity of the discussion, necessitating correspondingly nuanced adjustments from the AI. OPRO enables the model to respond effectively to such shifts, maintaining relevance and depth in its answers.

In this context, we introduce a novel prompt engineering approach tailored for cybersecurity threat modeling. Our method identifies potential threats, suggests mitigations, and references applicable security controls, such as those specified in NIST 800-53. Integrating the strengths of CoT and OPRO, our technique establishes a robust framework for threat modeling. By using CoT, we instruct LLMs to methodically outline their reasoning, simulating an expert’s analytical process in identifying and evaluating security threats. This clear reasoning is essential for validating AI-generated insights and ensures that each step is both logical and justifiable. The explicit articulation of thought processes not only deepens the model’s analytical capabilities but also enhances the reliability and traceability of its outputs, thereby enhancing domain knowledge application.

Alongside CoT, we employ OPRO to dynamically adjust these prompts according to the ongoing context of the threat modeling session. This integration allows our method to adaptively respond to new information or changes in focus, ensuring that the model’s analysis remains comprehensive and pertinent throughout the interaction. By merging these strategies, our approach guides LLMs to generate detailed, actionable threat models that not only identify potential risks but also recommend suitable mitigations aligned with NIST 800-53 control codes. This sophisticated prompting strategy significantly boosts the AI’s capacity to emulate expert-level cybersecurity analysis and aligns its outputs with industry standards, providing a formidable tool for organizations aiming to enhance their security measures.

\subsection{LLM Fine-tuning}
Fine-tuning is a crucial step in adapting pre-trained language models to specific tasks, enhancing their accuracy and effectiveness in specialized domains. In the context of cybersecurity, particularly for threat identification from Data Flow Diagrams (DFDs), fine-tuning enables the model to better understand domain-specific language patterns and structures associated with potential vulnerabilities. We employ Low-Rank Adaptation (LoRA) \cite{hu2021lora} as the primary technique for fine-tuning, which is both efficient and effective for resource-constrained environments.

\subsubsection{Low-Rank Adaptation (LoRA) for Fine-tuning}
LoRA is a parameter-efficient fine-tuning method that adapts large pre-trained models by injecting learnable low-rank matrices into the original model’s weights. This approach significantly reduces the number of trainable parameters, making fine-tuning more computationally feasible while maintaining model performance.

Let $\mathbf{W} \in \mathbb{R}^{d \times k}$ be the weight matrix of the pre-trained model, where $d$ is the input dimension and $k$ is the output dimension. In LoRA, the update to $\mathbf{W}$ is represented as the product of two low-rank matrices:
\[
\Delta \mathbf{W} = \mathbf{A} \mathbf{B}
\]
where $\mathbf{A} \in \mathbb{R}^{d \times r}$ and $\mathbf{B} \in \mathbb{R}^{r \times k}$, with $r \ll \min(d, k)$ being the rank of the decomposition. This decomposition allows the original weight matrix to be updated as:
\[
\mathbf{W}_{\text{new}} = \mathbf{W} + \alpha \cdot \Delta \mathbf{W}
\]
where $\alpha$ is a scaling factor to control the magnitude of the adaptation, and $\Delta \mathbf{W}$ is the low-rank update applied to the original weight matrix $\mathbf{W}$.

\textbf{Principles of LoRA Fine-tuning}
The core principles of the LoRA fine-tuning method include:

\begin{itemize}
    \item \textbf{Parameter Efficiency}: By training only the low-rank matrices $\mathbf{A}$ and $\mathbf{B}$, LoRA significantly reduces the number of trainable parameters. The total number of trainable parameters becomes $d \times r + r \times k$, where $r \ll \min(d, k)$, making it highly efficient compared to traditional fine-tuning methods.
    
    \item \textbf{Computational Efficiency}: Since only the low-rank matrices $\mathbf{A}$ and $\mathbf{B}$ are optimized during training, the computational and memory requirements are substantially lower than those of standard fine-tuning methods. This makes LoRA suitable for environments with limited computational resources, such as edge devices or lower-end GPUs.

    \item \textbf{Maintaining Performance}: Despite its reduced parameter count, LoRA maintains or even enhances the model's performance. The low-rank updates effectively capture domain-specific features without compromising the model’s generalization capability.

    \item \textbf{Adaptability to Domain-specific Tasks}: LoRA allows for efficient adaptation of pre-trained models to specialized tasks by focusing on learning task-specific information encoded within the low-rank matrices. In the context of threat modeling from DFDs, LoRA helps the model recognize patterns and relationships specific to cybersecurity.

\end{itemize}

The LoRA-based fine-tuning approach thus provides a robust framework for adapting pre-trained models to domain-specific tasks in a resource-efficient manner, making it well-suited for applications like automated threat modeling.






\section{Experimental Setting}

\subsection{Data Preparation}
For this study, we manually created 50  sample related to the banking system domain, each representing a different application.
Each sample simulates a unique banking system architecture, covering various aspects of threat modeling to ensure diversity and robustness in the training data. For each sample, it contains two fields: 
\begin{itemize}
    \item \textbf{Application Description}: Each sample contains a textual description of the application system, outlining its components, data flow, and overall architecture.
    \item \textbf{Ground Truth Threats and Mitigations}: Each application description includes manually verified threats and mitigations, which serve as ground truth labels for training and evaluation.
\end{itemize}

\subsection{Fine-tuning Configuration}
The fine-tuning process leverages LoRA to enhance the model’s performance in the banking threat modeling domain. Key parameters for LoRA include:

\begin{itemize}
    \item \textbf{Rank (r)}: 32, specifying the rank of the low-rank decomposition, balancing between parameter efficiency and model adaptation.
    \item \textbf{Scaling Factor (lora\_alpha)}: 64, controlling the adaptation strength of the model.
    \item \textbf{Target Modules}: Includes projections such as ``q\_proj'', ``k\_proj'', ``v\_proj'', and ``o\_proj'', which are specific to Llama models.
    \item \textbf{Dropout Rate}: 0.1, applied to prevent overfitting during the adaptation process.
\end{itemize}


\subsection{Training Configuration}
The model training was conducted using the Transformers library with the following major hyperparameters:

\begin{itemize}
    \item \textbf{Batch Size}: 4 per device, with gradient accumulation steps set to 4, effectively increasing the batch size during training.
    \item \textbf{Optimizer}: ``paged\_adamw\_32bit'', chosen for memory efficiency and faster convergence.
    \item \textbf{Learning Rate}: 1e-4, set for stable learning and effective adaptation.
    \item \textbf{Number of Epochs}: 30, allowing sufficient training for the model to adapt to the threat modeling domain.
    \item \textbf{Evaluation Strategy}: Evaluations are performed at regular intervals (every 20\% of training steps), ensuring consistent monitoring of the model's performance.
\end{itemize}

\begin{figure*}
    \centering
    \includegraphics[width=1.0\textwidth]{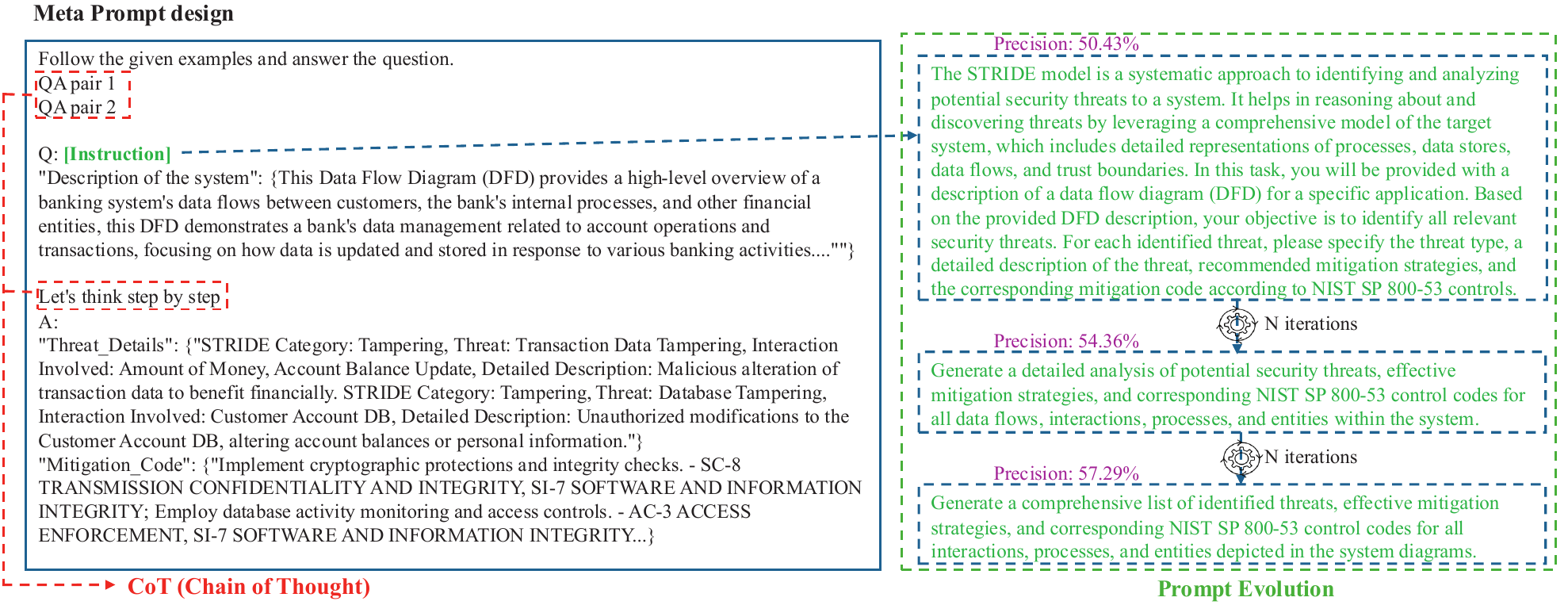}
    \caption{The process of prompt design, combing CoT and Prompt Evolution based on OPRP.}
    \label{fig:engineering_prompt_process}
\end{figure*}

\subsection{Initial Prompt}

For prompt engineering, we use the following initial prompt as a start point for instructing the LLMs.

\tcbset{colback=gray!5!white, colframe=gray!75!black, fonttitle=\bfseries}

\begin{tcolorbox}[title=Initial Prompt for Prompt Engineering, label=box:initial_prompt]
The STRIDE model is a systematic approach to identifying and analyzing potential 
security threats to a system. It helps in reasoning about and discovering threats 
by leveraging a comprehensive model of the target system, which includes detailed 
representations of processes, data stores, data flows, and trust boundaries. In this task, you will be provided with a description of a data flow diagram (DFD) 
for a specific application. Based on the provided DFD description, your objective 
is to identify all relevant security threats. For each identified threat, please 
specify the threat type, a detailed description of the threat, recommended mitigation 
strategies, and the corresponding mitigation code according to NIST SP 800-53 controls.
\end{tcolorbox}

\subsection{Prompt Configuration}
The prompt engineering process includes two steps: CoT (Chain of Thoughts) and prompt evolution as shown in Figure~\ref{fig:engineering_prompt_process}. For CoT, we incorporate ``few-shot'' learning
by utilizing two examples to guide prompt design and
optimization, and ``zero shot'' by employing a step-by-step reasoning
method to address specific threats within the banking
system.

The prompt evolution is implemented based on OPRO, with the following configurations:

\textbf{Scorer:}
\begin{itemize}
    \item \textbf{Model type}: ``gpt-3.5-turbo'' 
    \item \textbf{Max output tokens}: 1024. A higher token limit allows for comprehensive responses that can fully evaluate the effectiveness and completeness of the prompts.
    \item \textbf{temperature}: 0.0. The model will produce the most likely output, which is beneficial for consistent scoring and easier comparative analysis.
    \item \textbf{Num Decodes}: 1. Since the focus is on reliability and predictability for scoring, only one decode is needed to evaluate each input without introducing variability.
    \item \textbf{Batch size}: 1. Maintain simplicity and control over the experiment. Each prompt is processed individually, reducing complexity in handling outputs.
    \item \textbf{Num Servers}: 1. Simplify the infrastructure requirements and ensures that the environmental factors affecting model performance are consistent across all tests.
\end{itemize}

\textbf{Optimizer}:
\begin{itemize}
    \item \textbf{Model type}: ``gpt-3.5-turbo'' 
    \item \textbf{Max output tokens}: 512.  A lower token count encourages the model to focus on conciseness and creativity within a shorter output, which might lead to more diverse and inventive prompt generation.
    \item \textbf{temperature}: 1.0. Increase randomness and variability in responses. This setting is optimal for generating creative and diverse prompts.
    \item \textbf{Batch size}: 1. Like the scorer, maintaining a batch size of 1 ensures that each generated prompt is evaluated individually, allowing for precise adjustments and optimizations based on singular output analysis.
    \item \textbf{num\_servers}: 1. Consistency in computational environment between scoring and optimizing, reducing any potential bias or variability introduced by different server setups.
\end{itemize}

Other settings:
\begin{itemize}
    \item \textbf{Instruction position}: ``Q\_begin'' (the instruction is added before the original question.)
\end{itemize}

\subsection{Evaluation Metrics}

To evaluate the performance of the proposed prompt engineering and the fine-tuned methods to identify the threats and proper mitigation control codes,  we adopted semantic similarity analysis using BERT and set-based evaluation of mitigation control codes using precision, recall, and accuracy.

\textbf{BERT Similarity Score:} We used the BERT model to compute the cosine similarity between the embeddings of the LLM-generated threats/mitigation strategies and the ground truth annotations. This allows us to capture not only exact matches but also paraphrased or contextually similar descriptions.

The similarity score for each generated output is calculated as:
\[
\text{Similarity Score} = \cos(\text{BERT}(G_{\text{truth}}), \text{BERT}(G_{\text{generated}}))
\]
where \( G_{\text{truth}} \) is the ground truth description, and \( G_{\text{generated}} \) is the LLM-generated output.

Higher similarity scores indicate better alignment with the ground truth. 

\subsubsection{Mitigation Control Code Evaluation}

To evaluate the accuracy of the mitigation control codes generated by the LLM, we treated this as a set-matching problem, comparing the sets of control codes from the generated output to the ground truth. We employed the following metrics to assess performance:

\textbf{Precision:} Precision measures the proportion of correctly generated control codes out of all codes predicted by the LLM. It is calculated as:
\[
\text{Precision} = \frac{|C_{\text{generated}} \cap C_{\text{truth}}|}{|C_{\text{generated}}|}
\]
where \( C_{\text{generated}} \) represents the set of control codes generated by the LLM, and \( C_{\text{truth}} \) represents the set of ground truth control codes.

\textbf{Recall:} Recall measures the proportion of the correct control codes identified out of all the ground truth codes. It is calculated as:
\[
\text{Recall} = \frac{|C_{\text{generated}} \cap C_{\text{truth}}|}{|C_{\text{truth}}|}
\]

\textbf{Accuracy:} Accuracy evaluates the overall correctness of the generated control codes, comparing the total number of correctly generated codes to the total number of codes present in both the generated set and the ground truth set. It is calculated as:
\[
\text{Accuracy} = \frac{|C_{\text{generated}} \cap C_{\text{truth}}|}{|C_{\text{generated}} \cup C_{\text{truth}}|}
\]

The  $C_{\text{generated}}$ and $C_{\text{truth}}$ denote the mitigation codes generated by LLM and the ground truth mitigation codes. 

\section{Evaluation}

In this study, we aim to address the following research questions:

\textbf{RQ1: How does the performance of CoT+OPRO compare to the Initial Prompt, CoT, and OPRO?}

\textbf{RQ2: How does fine-tuning improve the performance of LLMs compared to their base models?}

\textbf{RQ3: How does the performance of our developed system, ThreatModeling-LLM, compare to existing methodologies? What are the effects of integrating prompt engineering with fine-tuning within our ThreatModeling-LLM framework?}

In the following sections, we answer the research questions and present our empirical findings.



\section*{RQ1: How does the performance of CoT+OPRO compare to the Initial Prompt, CoT, and OPRO?}

Figure~\ref{fig:engineering_prompt_process} highlights two crucial components in our meta prompt design process:  CoT and Prompt Evolution based on OPRO. These elements play a central role in refining prompt design to enhance threat identification and mitigation precision within a banking data flow diagram scenario.
Starting with CoT, it employs a step-by-step reasoning method to address specific threats within the banking system, such as data tampering and unauthorized modifications. This zero-shot approach enables targeted solutions directly linked to identified risks, ensuring that each threat is systematically addressed with appropriate security controls. In addition, CoT incorporates “few-shot” learning by utilizing specific examples to guide prompt design and optimization. This technique leverages two example QA pairs, representing distinct scenarios or questions related to potential security vulnerabilities, to refine the analysis and understanding of security threats in banking data flow diagrams.

The Prompt Evolution segment captures how iterative refinements based on OPRO significantly enhance output precision. Initial analysis precision is at 0.5, reflecting the early stage of addressing security vulnerabilities. Continuous iterations refine threat models, improve description clarity, and optimize mitigation strategies, progressively increasing precision. This process culminates in a final precision of 0.57, demonstrating a systematic approach to maximizing the effectiveness of prompt design for identifying and mitigating potential security threats in the banking context.

\begin{figure}
    \centering
    \includegraphics[width=0.5\textwidth]{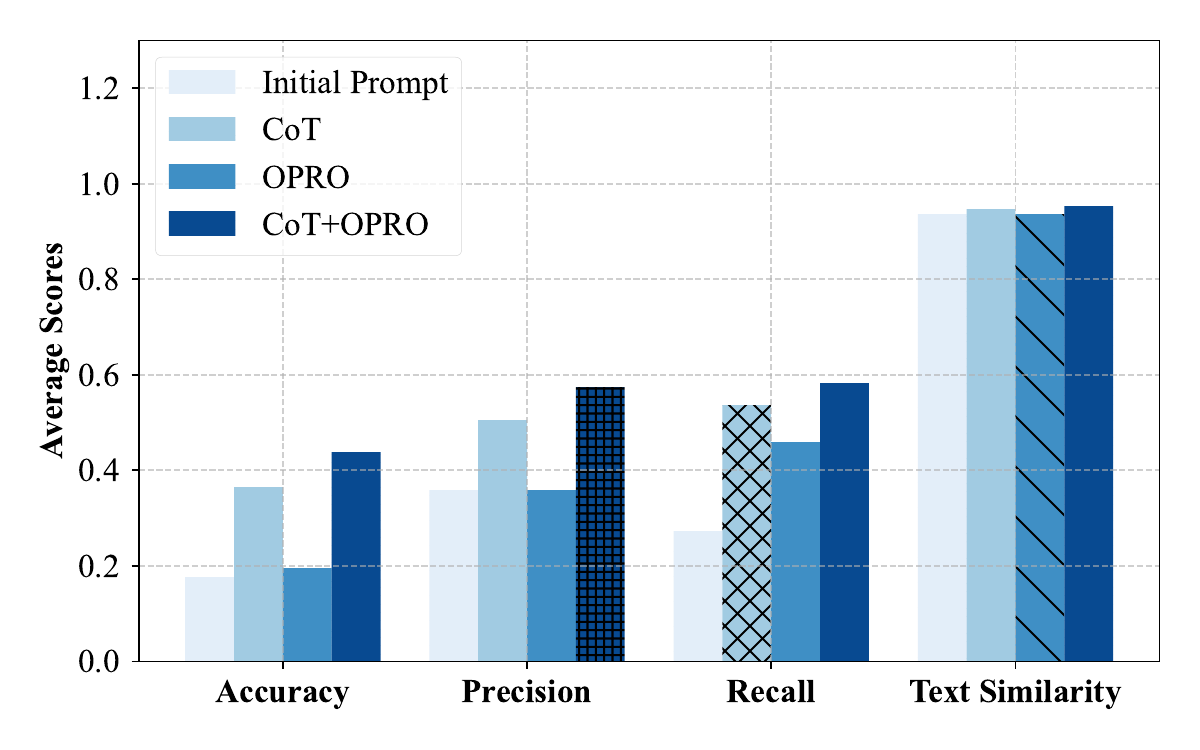}
    \caption{Comparative performance analysis of four prompt engineering techniques (Initial Prompt, CoT, OPRO, and the combined CoT+OPRO), across Accuracy, Precision, Recall, and Text Similarity metrics based on GPT-3.5-turbo. 
    }
    \label{fig:engineering_prompt_result}
\end{figure}

Figure~\ref{fig:engineering_prompt_result} presents a comparative analysis of four prompt engineering techniques: Initial Prompt, CoT, OPRO, and the combined CoT+OPRO approach. Evaluated across four metrics (Accuracy, Precision, Recall, and Text Similarity), the results, based on GPT-3.5-turbo, reveal that integrating CoT with OPRO achieves the highest performance across all metrics, particularly excelling in Precision and Text Similarity.

The Initial Prompt method, serving as a baseline, shows limited effectiveness, with relatively lower scores in Accuracy (0.17), Precision (0.35), and Recall (0.27). CoT improves upon this baseline by significantly increasing both Accuracy and Recall, demonstrating that explicit reasoning steps contribute to better threat identification. CoT’s Precision also surpasses the baseline, highlighting its capability to generate more relevant and precise outputs. OPRO, when applied independently, achieves moderate improvements, especially in Precision, but does not match CoT’s overall performance.

The combination of CoT and OPRO demonstrates a synergistic effect, achieving the highest scores in all metrics, with Precision and Recall almost achieving 0.6, and Text Similarity exceeding 0.95. This combined approach enables the model to reason through intermediate steps (CoT) while dynamically refining prompts (OPRO), producing more accurate and contextually relevant outputs. These results indicate that CoT+OPRO not only enhances overall performance but also addresses the limitations of each individual method, making it a robust strategy for improving prompt engineering in automated threat modeling tasks.
The output of prompt engineering process will lead to a new prompt as shown below: 

\tcbset{colback=gray!5!white, colframe=gray!75!black, fonttitle=\bfseries}

\begin{tcolorbox}
[title=Optimized Prompt by the proposed Prompt Engineering method,label=box:new_prompt]
Generate a comprehensive list of identified threats, effective mitigation strategies, and corresponding NIST SP 800-53 control codes for all interactions, processes, and entities depicted in the system diagrams.
\end{tcolorbox}

We began the prompt engineering process with an initial prompt shown in section  \ref{box:initial_prompt} that was detailed and descriptive, aiming to guide the model through a comprehensive analysis of threats based on a given Data Flow Diagram (DFD). This prompt provided explicit instructions for identifying security threats, specifying their types, and recommending mitigation strategies along with corresponding NIST SP 800-53 controls. Through iterative refinement, the prompt was optimized to a more compact version that retains essential information while enhancing clarity and efficiency. The optimized prompt focuses on generating a comprehensive analysis of threats, mitigation strategies, and relevant NIST SP 800-53 controls for identified risks. This evolution in prompt design reflects a balance between thorough guidance and streamlined communication, improving the model’s response precision.


\begin{figure}[ht]
    \centering
    \begin{subfigure}{0.5\textwidth}
        \centering
        \includegraphics[width=\linewidth]{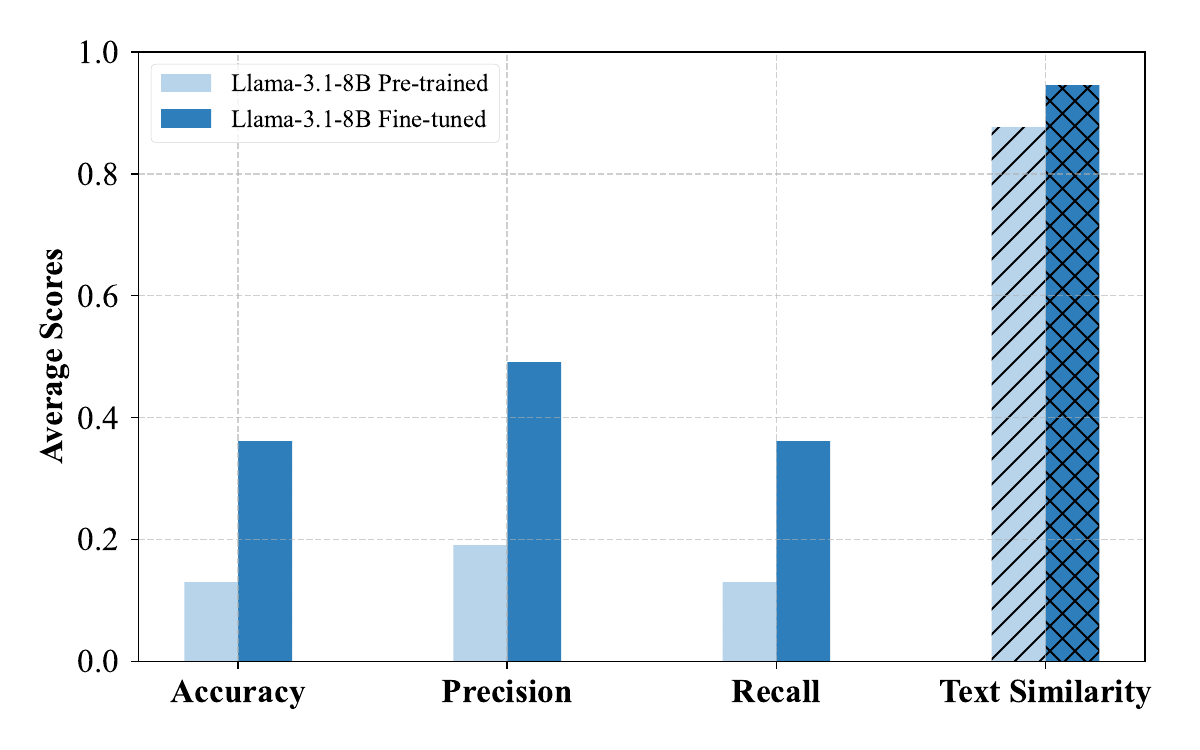}
        \caption{Llama-3.1-8B}
        \label{fig:finetune_Llama-3.1-8B}
    \end{subfigure}

    \vspace{0.5cm} 

    \begin{subfigure}{0.5\textwidth}
        \centering
        \includegraphics[width=\linewidth]{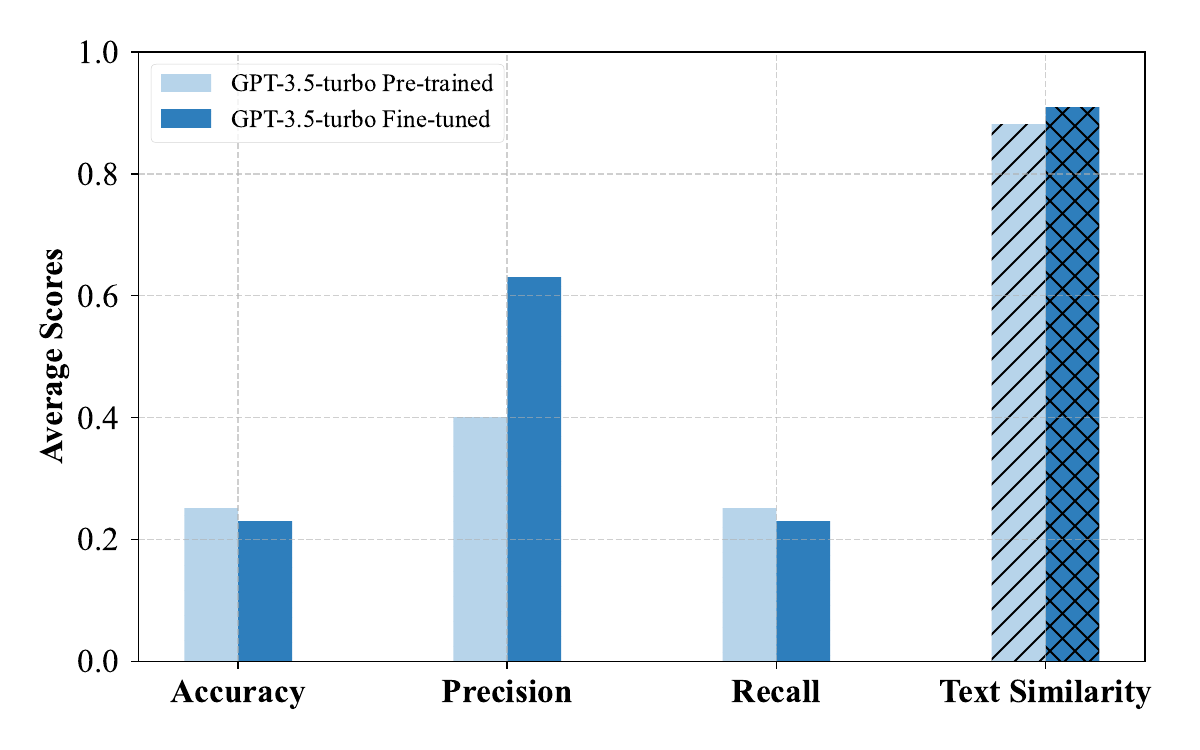}
        \caption{GPT-3.5-turbo}
        \label{fig:finetune_GPT-3.5-turbo}
    \end{subfigure}

    \caption{Comparison of Precision, Recall, Accuracy, and Text Similarity scores for Llama-3.1-8B and GPT-3.5-turbo:
base models versus fine-tuned models. The results indicate significant performance improvements across all metrics
with fine-tuning.}
    \label{fig:LLM_base_finetune}
\end{figure}

\begin{tcolorbox}[colback=gray!20,boxsep=3pt,left=1pt,right=1pt,top=1pt,bottom=1pt]
\noindent \textbf{Takeaway 1}: Combining Chain of Thought (CoT) with Optimization by PROmpting (OPRO) results in the highest accuracy, precision, recall and text similarity. Performance metrics showed an increase from 0.35 in initial prompt precision to 0.57 in CoT+OPRO.
\end{tcolorbox}

\section*{RQ2: How does fine-tuning improve the performance of LLMs compared to their base models?}

Figure~\ref{fig:LLM_base_finetune} presents a comparative analysis of the performance between base and fine-tuned models for Llama-3.1-8B (\ref{fig:finetune_Llama-3.1-8B}) and GPT-3.5-turbo (\ref{fig:finetune_GPT-3.5-turbo}). The evaluation covers four key metrics: Accuracy, Precision, Recall, and Text Similarity. The results demonstrate that fine-tuning significantly enhances model performance across all metrics, with the most substantial improvements observed in Precision and Recall.

The base models exhibit moderate performance, with Accuracy and Recall generally lower than 0.3. Fine-tuning, however, substantially boosts these metrics. For instance, the fine-tuned version of Llama-3.1-8B achieves an Accuracy score above 0.6 and a Recall score nearing 0.5, indicating better identification of relevant threats and generation of mitigation strategies. Similarly, GPT-3.5-turbo, while having moderate base performance, shows significant improvements after fine-tuning, particularly in Precision, which increases from approximately 0.4 to 0.6. This demonstrates the model's enhanced ability to generate more relevant outputs and reduce false positives.

Moreover, the results consistently reveal improvements in Text Similarity, with fine-tuned models generating outputs that align more closely with ground truth annotations. Fine-tuned Llama-3.1-8B achieves nearly perfect Text Similarity, highlighting the effectiveness of fine-tuning in adapting language models for domain-specific tasks like automated threat modeling.

It is important to note that, due to limited GPU resources, larger open-source models, such as Llama-3.1-70B, were not fine-tuned in this study. The computational demands of fine-tuning these larger models exceeded available resources. As a result, the analysis was focused on the smaller, more feasible models, demonstrating that even with resource constraints, fine-tuning smaller models like Llama-3.1-8B and GPT-3.5-turbo can yield significant performance gains.

\begin{tcolorbox}[colback=gray!20,boxsep=3pt,left=1pt,right=1pt,top=1pt,bottom=1pt]
\noindent \textbf{Takeaway 2}: Fine-tuning LLMs enhances their ability to identify and mitigate threats significantly, improving precision from 0.4 to 0.6 and recall from 0.3 to 0.5 in banking threat modeling tasks.
\end{tcolorbox}

\section*{RQ3: How does the performance of our developed system, ThreatModeling-LLM, compare to existing methodologies? What are the effects of integrating prompt engineering with fine-tuning within our ThreatModeling-LLM framework?}

\begin{figure}[!t]
    \centering
    \includegraphics[width=1\linewidth]{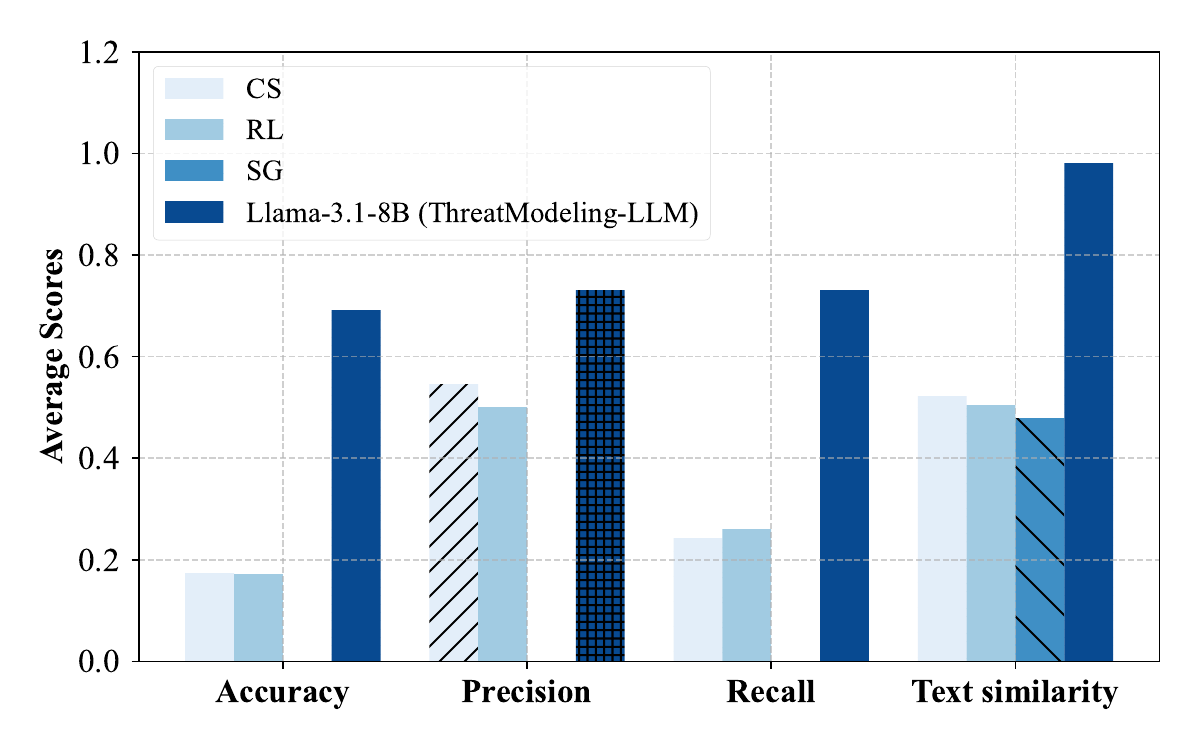}

    \caption{Performance comparison of GPT-based threat modelling tools and our method (ThreatModeling-LLM). This figure presents comparative results across our method and three baseline tools: STRIDEGPT (SG), ChatGPT as a raw LLM (RL), and Cyber Sentinel (CS). The metrics evaluated include text similarity (based on Bert) and effectiveness in mapping to NIST control codes, measured in terms of accuracy, precision, and recall.}
    \label{fig:baseline_comparison}
  
\end{figure}

We firstly compared the performance of ThreatModeling-LLM (applied on the pre-trained Llama3.1-8B) to the three baselines models which used emergent GPT-based technologies, as detailed in Section~\ref{sec:motivation}.
These tools are Cyber Sentinel (CS)~\cite{kaheh2023cyber}, STRIDEGPT (SG)~\cite{stride_gpt}, and Raw LLM (RL) (using ChatGPT)~\cite{chatgpt} stand out for their unique capabilities.
Figure~\ref{fig:baseline_comparison} illustrates the comparative analysis between ThreatModeling-LLM and the traditional baselines.
We excluded pytm for comparison as it cannot generate NIST 800-53 control codes, hence there are no results according to our performance metrics. 
SG, in particular, showed no results on the first metric of mapping NIST 800-53 control codes due to its inability to generate these codes. 
The results reveal significant limitations across all tools, SG, RL, and CS, in their capability to accurately map NIST 800-53 control codes and produce textually aligned threats and mitigations. 
For instance, all tools showed a notable gap in precision, scoring below 0.6, and struggled with a recall below 0.30. 
The text similarity results further highlight a gap in all tools’ ability to produce text outputs that closely match the semantic and syntactic requirements of the control codes and ground truth. In stark contrast, ThreatModeling-LLM  outperformed these metrics significantly, achieving precision and recall rates exceeding 0.70, illustrating its superior capability in aligning with NIST 800-53 compliance standards and delivering more accurate and reliable threat modeling outcomes.

\begin{figure}[!t]
    \centering
    \begin{subfigure}{0.5\textwidth}
        \centering
        \includegraphics[width=\linewidth]{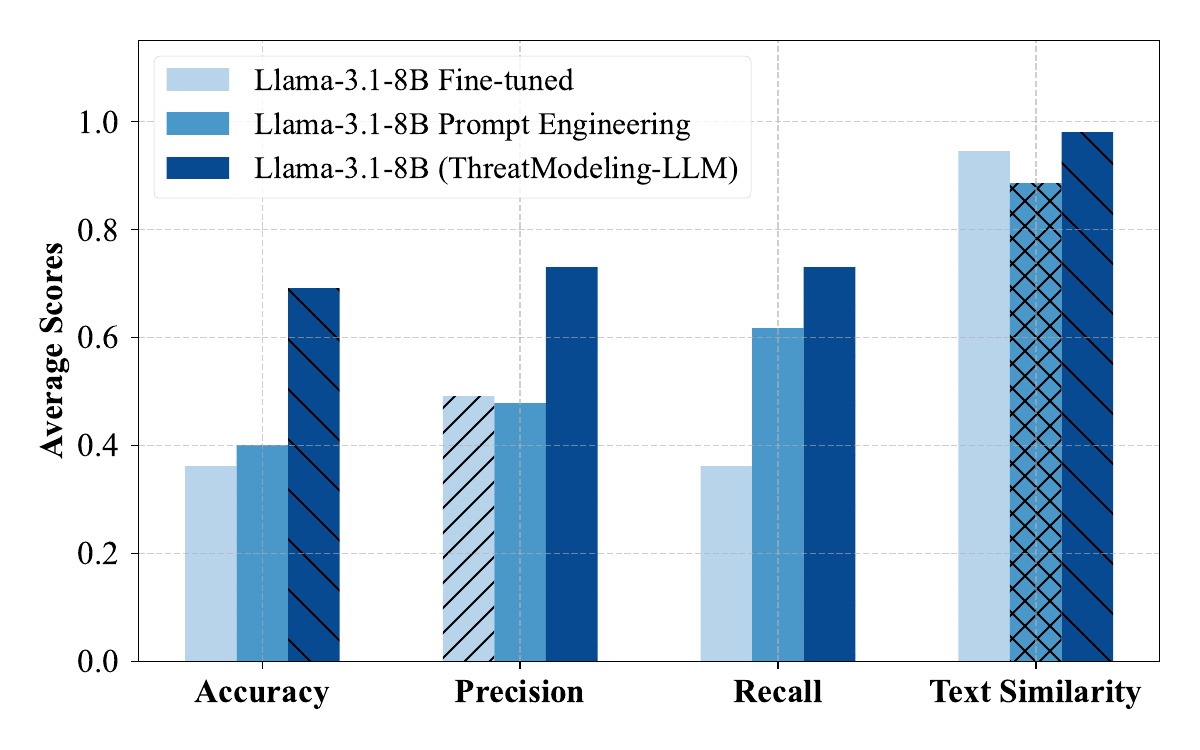}
        \caption{Llama-3.1-8B}
        \label{fig:prompt_finetune_Llama-3.1-8B}
    \end{subfigure}

    \vspace{0.5cm} 

    \begin{subfigure}{0.5\textwidth}
        \centering
        \includegraphics[width=\linewidth]{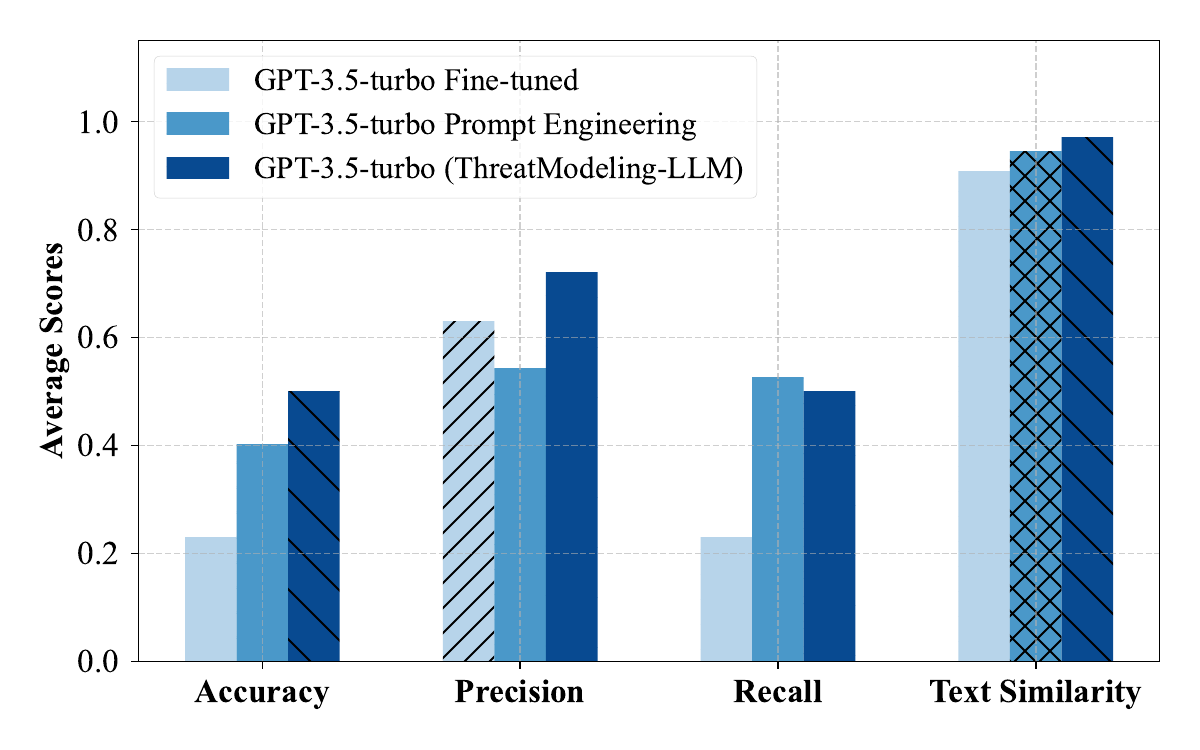}
        \caption{GPT-3.5-turbo}
    \label{fig:prompt_finetune_GPT-3.5-turbo}
    \end{subfigure}

    \caption{Impact of combining prompt engineering with fine-tuning on the performance of LLMs (Llama-3.1-8B and GPT-3.5-turbo) in threat modeling tasks. Llama-3.1-8B shows a dramatic improvement across all metrics, particularly in Accuracy, Precision, and Recall, when prompt engineering is combined with fine-tuning.}
    \label{fig:prompt_engineering_finetune}
\end{figure}

We further investigate  the effects of integrating prompt engineering with fine-tuning within our ThreatModeling-LLM framework. 
Figure~\ref{fig:prompt_engineering_finetune} compares the performance of Llama-3.1-8B (\ref{fig:prompt_finetune_Llama-3.1-8B}) and GPT-3.5-turbo (\ref{fig:prompt_finetune_GPT-3.5-turbo}) across four metrics—Accuracy, Precision, Recall, and Text Similarity—under different conditions: fine-tuned models, models with prompt engineering, and models applied with ThreatModeling-LLM. The results clearly show that the combination of prompt engineering and fine-tuning achieves the highest performance across all metrics, with Llama-3.1-8B exhibiting the most substantial improvement.

For Llama-3.1-8B, fine-tuning alone achieves an accuracy of 0.36, precision of 0.49, recall of 0.36, and text similarity of 0.944. Prompt engineering alone results in modest improvements, pushing accuracy to 0.4001, precision to 0.4773, and recall to 0.6162. However, when applied with the ThreatModeling-LLM, Llama-3.1-8B shows a dramatic boost, achieving an accuracy of 0.69, precision of 0.73, recall of 0.73, and text similarity of 0.9792. This substantial increase across all metrics highlights how prompt engineering complements the model’s fine-tuned understanding, enabling more precise and comprehensive identification of threats and mitigations.

For GPT-3.5-turbo, fine-tuning alone yields an accuracy of 0.23, precision of 0.63, recall of 0.23, and text similarity of 0.908. Prompt engineering without fine-tuning produces higher scores, reaching 0.4019 in accuracy, 0.5421 in precision, 0.5251 in recall, and 0.9449 in text similarity. The combination of prompt engineering and fine-tuning achieves further gains, with an accuracy of 0.50, precision of 0.72, recall of 0.50, and text similarity of 0.9710. While the improvements are notable, they are more moderate compared to Llama-3.1-8B, suggesting that the combined approach has a stronger impact on Llama-3.1-8B’s performance.

Overall, these results confirm that integrating prompt engineering with fine-tuning maximizes model performance, particularly for Llama-3.1-8B. The combined strategy not only enhances accuracy and recall but also significantly improves precision and text similarity, leading to outputs that are more relevant and contextually aligned with ground truth annotations. This demonstrates the effectiveness of using both methods together to improve automated threat modeling.

\begin{tcolorbox}[colback=gray!20,boxsep=3pt,left=1pt,right=1pt,top=1pt,bottom=1pt]
\noindent \textbf{Takeaway 3}: Our analysis shows that ThreatModeling-GPT significantly outperforms all baseline models, including Cyber Sentinel, STRIDEGPT, and Raw LLM (ChatGPT). The integrated approach of prompt engineering with fine-tuning shows the greatest improvement, elevating precision from 0.49 to 0.73 and accuracy from 0.36 to 0.69.
\end{tcolorbox}

\section{Discussion}

ThreatModeling-LLM effectively automates threat modeling across various pre-trained LLMs, demonstrating adaptability and improved compliance with NIST 800-53 controls. Fine-tuning models like GPT-3.5 and Llama-3.1 enhances their accuracy in threat identification and mitigation, while techniques like CoT prompting and OPRO optimization further boost performance. Although the current focus is on banking systems, future research could extend this approach to sectors like healthcare and IoT. Addressing challenges such as generalization across domains and optimizing resource efficiency for larger models remains crucial for broader application.

This study mainly focuses on LLM-based automatic threat modeling using 50 samples. However, it is worth noting that our dataset is unique, as it was carefully collected in collaboration with a local bank. The data is not only rare but has been thoroughly verified and is crucial for research in this area. Expert verification and GPT augmentation have been used to enhance the dataset’s representativeness. Our threat modeling for banking systems is different from other sectors, like the stock market. The stock market deals with time-series data that changes constantly, while our data reflects real scenarios banks face. Since banks use a limited number of software applications, our dataset already covers most of the key situations they encounter. In the future, we plan to expand this approach to other sectors, such as the stock market, where we can apply it to large-scale data and solve real-world problems.

\textbf{Banking Sector Focus}: The study primarily addresses banking-related threats, limiting its current applicability. Future work will test generalizability to other sectors without compromising domain-specific precision.

\textbf{Implications}:
1) Cross-Domain Potential: While initially designed for banking, ThreatModeling-LLM is adaptable to various sectors through dataset customization and model retraining, making it a versatile cybersecurity tool.
2) Reduced Human Effort: The approach automates threat modeling, minimizing human intervention, reducing errors, and accelerating response times, making it scalable and resource-efficient for complex systems.

\section{Conclusion}
The role of Large Language Models  in cybersecurity, particularly in automating tasks like threat modeling, has demonstrated significant potential but remains underexplored. Our research illustrates that with proper prompt engineering and fine-tuning, LLMs can effectively automate threat modeling for banking systems, resulting in substantial improvements in both threat identification accuracy and the quality of mitigation strategies. By integrating techniques such as Chain of Thought  and Optimization by PROmpting alongside fine-tuning, our proposed system ThreatModeling-LLM achieves superior performance in detecting and addressing security vulnerabilities.
The results also emphasize that smaller, fine-tuned models like Llama-3.1-8B, when combined with prompt engineering, can dramatically enhance performance, even outperforming models like  GPT-3.5-turbo in key metrics. This makes them a resource-efficient solution that does not compromise accuracy, making Llama-3.1-8B particularly promising for real-world applications where computational resources are limited.

\section*{Acknowledgement}
The work has been supported by the Cyber Security Research Centre Limited whose activities are partially funded by the Australian Government’s Cooperative Research Centres Programme.

\newpage 

\bibliographystyle{IEEEtran}
\balance
\bibliography{reference}

\appendices

\end{document}